\documentclass[%
 reprint,
bibnotes,
 amsmath,amssymb,
 aps,
floatfix,
]{revtex4-2}

\usepackage{graphicx}
\usepackage{braket}
\usepackage{xcolor}
\usepackage{dcolumn}
\usepackage{bm}

\usepackage{array}
\usepackage{booktabs}
\usepackage{graphicx}
\usepackage{algorithm2e}

\usepackage{lipsum}

\begin{document}

\preprint{APS/123-QED}

\title{Quantum Theory of Optical Spin Texture in Chiral Tellurium Lattice}

\author{Pronoy Das$^{1}$}
  
\author{Sathwik Bharadwaj$^{1}$}
 
\author{Jungho Mun$^{2}$}%
\author{Xueji Wang$^{1}$}
\author{Junsuk Rho$^{2,3}$}

\author{Zubin Jacob$^{1}$}%
 \email{zjacob@purdue.edu}
 
\affiliation{
 $^{1}$Elmore Family School of Electrical and Computer Engineering, Birck Nanotechnology Center, and Purdue Quantum Science and Engineering Institute, Purdue University, West Lafayette, IN 47907, USA \\
 $^2$POSCO-POSTECH-RIST Convergence Research Center for Flat Optics and Metaphotonics, Pohang 37673, Republic of Korea \\
 $^3$Department of Mechanical Engineering, Department of Chemical Engineering, and Department of Electrical Engineering, Pohang University of Science and Technology (POSTECH), Pohang 37673, Republic of Korea
}

\date{\today}

\begin{abstract}
The absence of inversion symmetry in chiral tellurium (Te) creates unique spin textures within its electron waves. However, understanding textured optical waves within Te remains a challenge due to the semi-classical limitations of the long-wavelength approximation. To unveil textured optical eigenwaves in Te, we develop a spin-resolved deep-microscopic optical bandstructure, analogous to its electronic counterpart. We demonstrate that the degeneracies in this optical bandstructure is lifted by the twisted lattice of Te, which induces optical gyrotropy. Our theory shows excellent agreement with experimental optical gyrotropy measurements. At the lattice level, we reveal that the chirality of Te manifests as deep-microscopic optical spin texture within the optical wave. Our framework uncovers the finite-momentum origin of optical activity and provides a microscopic basis for light-matter interactions in chiral crystalline materials.
\end{abstract}

\maketitle

\textit{Introduction--} The electronic bandstructure encapsulates the spectrum of allowed electron waves within a crystal. This bandstructure is governed by the underlying crystal symmetries. For example, the lack of inversion symmetry results in lifting the spin degeneracy of electronic bands in natural chiral materials such as Te \cite{Geldiyev2023,Kawano2019}. This degeneracy lifting results in unique electronic spin textures within the electron waves of the crystal. Consequently, a compelling question arises regarding the nature of analogous wave phenomena in the optical domain. Specifically, we ask whether such optical eigenwaves with spin textures can manifest within the crystal lattice of natural chiral materials. The concept of optical spin textures is well-established in free space optics \cite{Yang2022,Yang2021,Kalhor2021} and in artificially engineered chiral structures \cite{Karakhanyan2024,Lin2024}, which characterizes the local optical polarization at sub-wavelength volumes. However, the existence of such optical spin textures has not been studied in natural materials.

In this Letter, we directly address this unexplored domain by unveiling the deep-microscopic optical spin texture inherent to the chiral Te crystal. To achieve this goal, we develop a deep microscopic optical theory of Te, which reveals the spin-resolved optical bandstructure of Tellurium's twisted lattice. This concept of optical bandstructure which we introduce is distinct from metamaterial and photonic crystal bandstructure since our work necessarily requires a quantum approach including crystalline symmetries. The deep microscopic bandstructure serves as a critical tool for identifying the hidden optical eigenwaves and their corresponding degree of circular polarization (DoCP) within the Te unit cell. 

\begin{figure}[h!tb]
\centering
\includegraphics[width=0.48\textwidth]{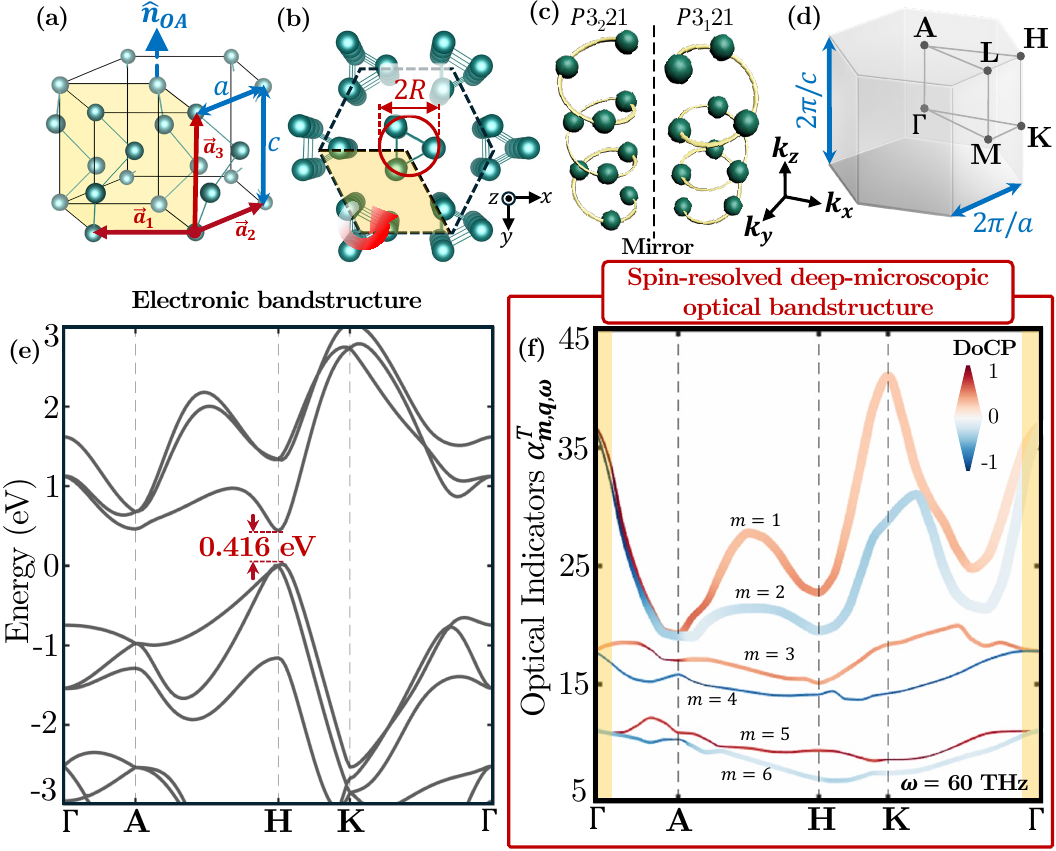}
\caption{\label{fig1} Spin-resolved deep-microscopic optical bandstructure of Te. (a) Schematic of the Te lattice, $\hat{n}_\text{OA}$ represents the screw axis. Here, $a=4.457 \mathring{A}$, $c=5.929 \mathring{A}$. (b) Top view of the Te crystal lattice. The unit cell is highlighted in yellow. Here, $2R=2.346 \mathring{A}$. (c) Screw orientation of the two enantiomers of chiral Te. (d) First Brillouin zone of Te crystal along with the high-symmetry points. (e)  Electronic band-structure of Te calculated using Purdue-PicoMax \textcolor{black}{(f) Spin-resolved deep-microscopic optical bandstructure  at $60$THz (=$5\mu$m). The classical limit of the bandstructure is highlighted in yellow near the $\mathbf{\Gamma}$ points. The dominant bands ($m=1,2$) determine the macroscopically measurable quantities such as the optical gyrotropy.}
}
\end{figure}

\begin{figure}[]
\centering
\includegraphics[width=0.48\textwidth]{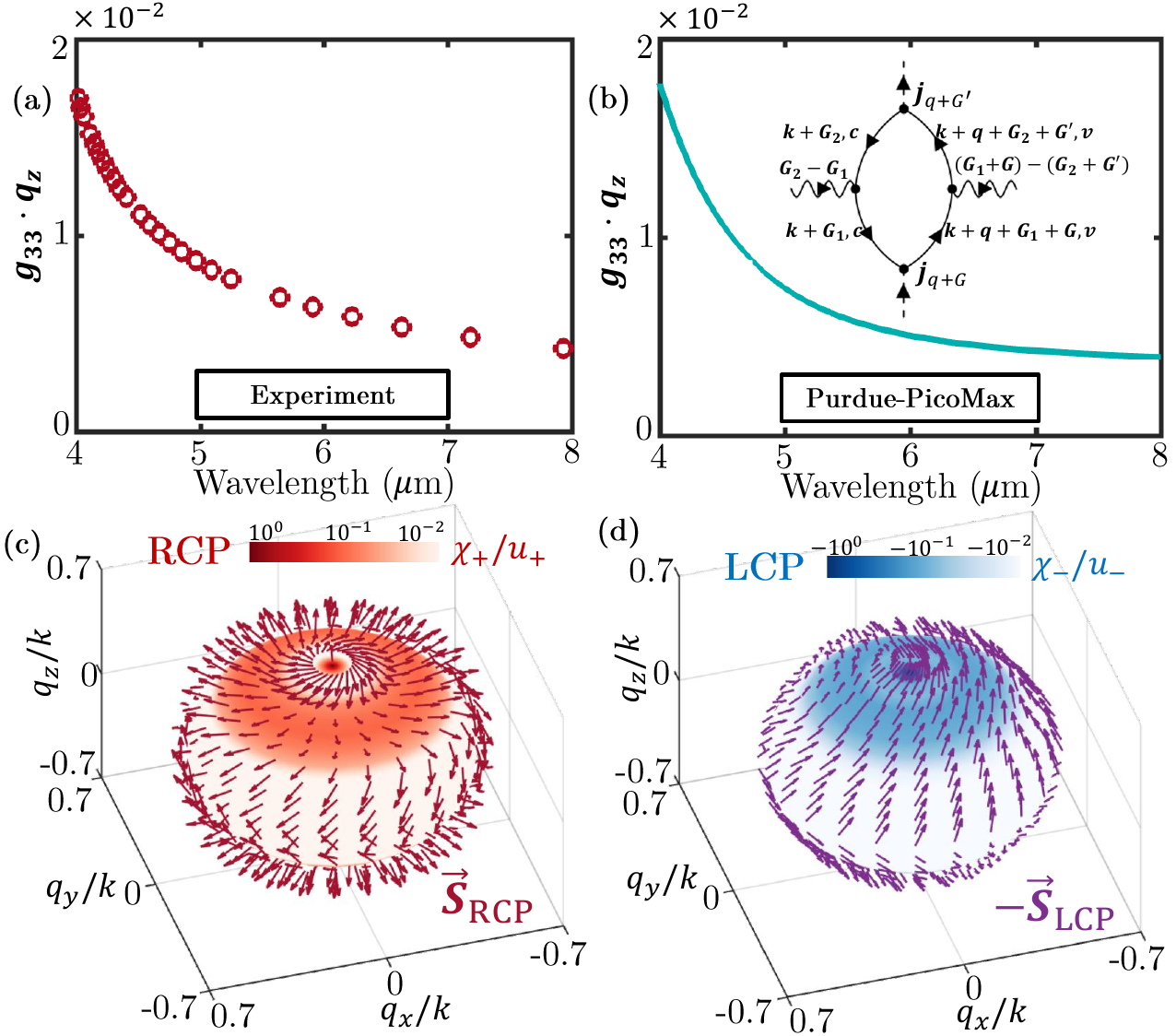}
\caption{\label{fig4} Super-dispersive optical gyrotropy and chirality density for undoped Te. (a) Experimental plot of the projection of the longitudinal gyration tensor component $\bm g_{33}$ on the photon momentum, vs. (b) Theoretical prediction using Purdue-PicoMax at $\bm{q}=(0,0,0.1)2\pi/a$ . The Feynman diagram in the inset reflects the transverse SILC of the induced transverse current including the momentum-exchange processes. Here, $c$ and $v$ represent the conduction and valence bands respectively. (c,d) The spin-resolved chirality density projection and total spin on the isofrequency surface of RCP ($+$) at $\omega=60$THz (c) and LCP ($-$) (d). The concentration near the pole of the sphere and the difference between RCP and LCP illustrates the suitable axis (optic axis) for observing high optical activity.
}
\end{figure}

\begin{figure*}[]
\centering
\includegraphics[width=0.55\textwidth]{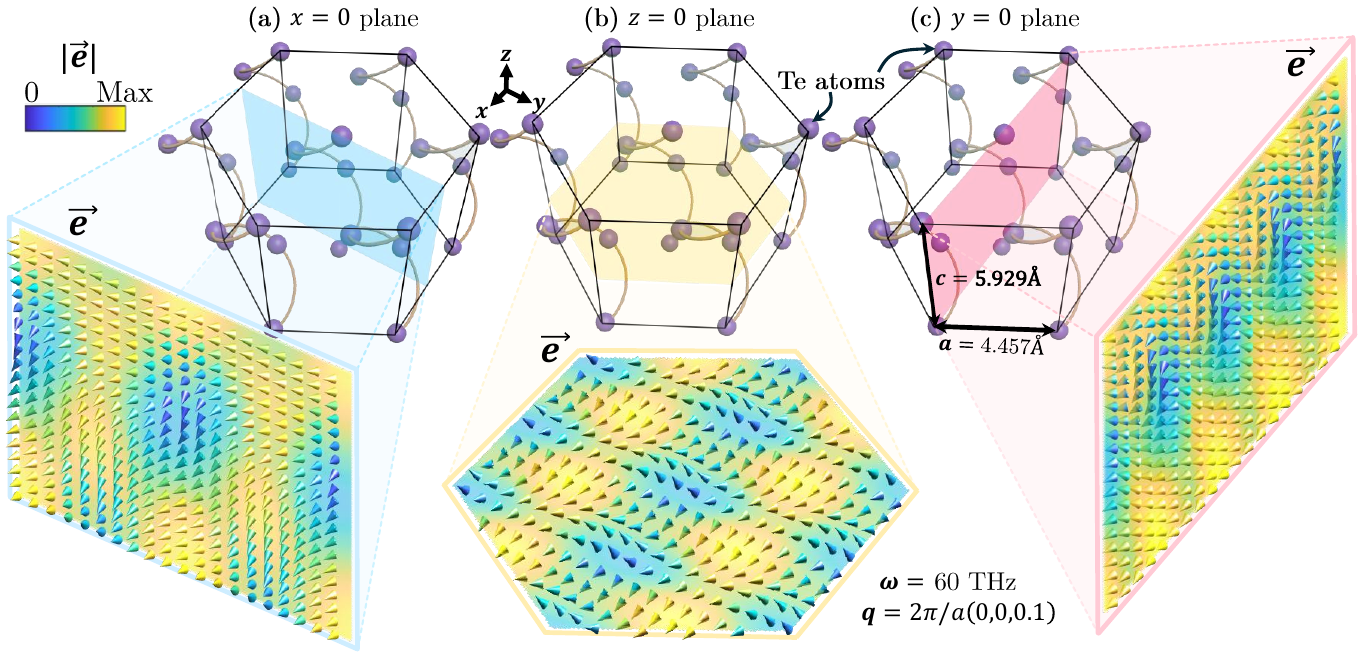}
\caption{\label{fig2} Hidden optical waves reveal polarization texture at the lattice-level. Vectorial distribution of the optical waves for the first transverse optical band. The plots are shown for three different planes of the Te crystal: (a) $x=0$, (b) $z=0$, and (c) $y=0$.
}
\end{figure*}
\begin{figure*}[]
\centering
\includegraphics[width=0.55\textwidth]{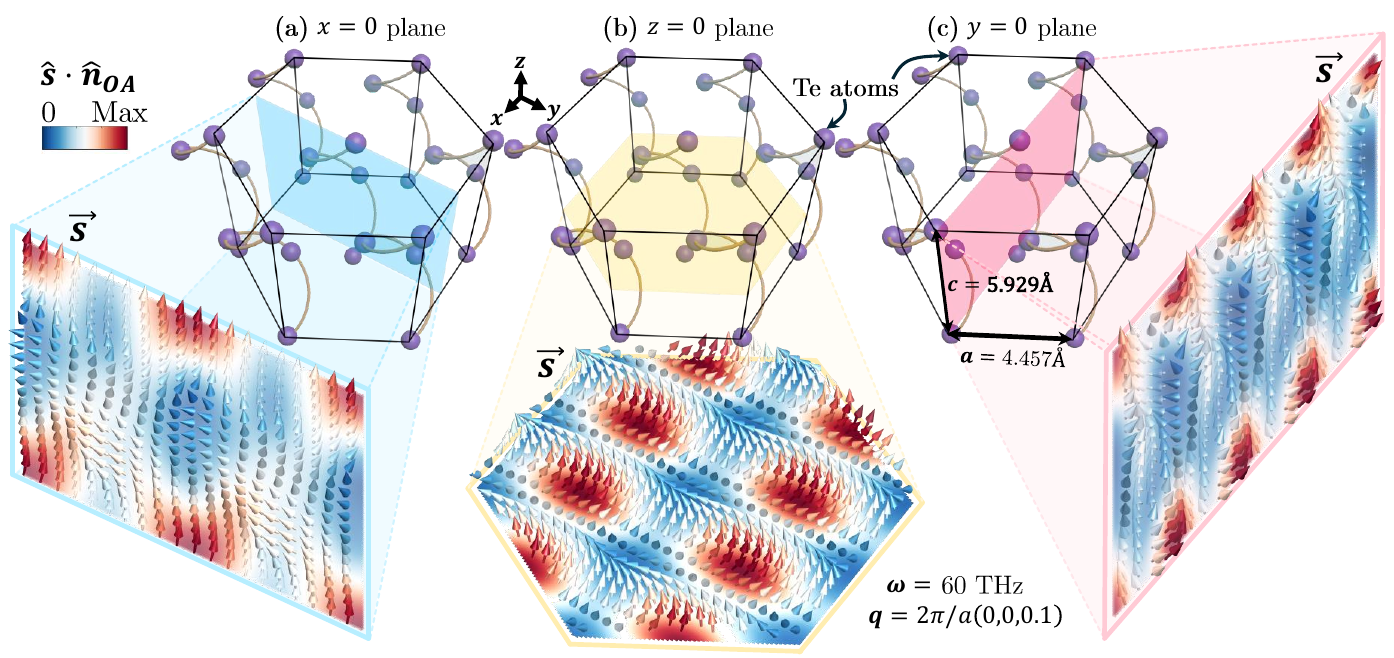}
\caption{\label{fig3} Deep-microscopic optical spin texture reveals orientation of polarization of the optical waves in Figure 3. Vectorial distribution of the optical spin texture of the optical waves in the first transverse optical band. The colormap shows the projection of the spin density along the screw axis. The plots are shown for three different planes of the Te crystal: (a) $x=0$, (b) $z=0$, and (c) $y=0$.
}
\end{figure*}

Our work reveals that the lack of inversion symmetry in Te lifts the degeneracy in the optical bandstructure, giving rise to optical gyrotropy. The deep-microscopic optical bandstructure captures the super-dispersive nature of this optical gyrotropy, and shows excellent agreement with experimental observations. We have implemented the necessary tools for calculating this optical gyrotropy within our open-source software: Purdue-PicoMax \cite{Bharadwaj2024}. Previously, there have been attempts to compute the optical gyrotropy through first-principles calculations \cite{Jnsson1996,Foster2010,Tsirkin2018,Wang2023,Zabalo2023}. However, they do not quantitatively capture the experimentally measured dispersion of optical gyrotropy across all wavelengths. This inconsistency stems from using the long-wavelength (near-zero photon momentum) expansion of the electromagnetic response function, whereas the optical gyrotropy is fundamentally a finite-momentum effect (i.s. spatial dispersion of matter). Here, we resolve this long-standing problem by accurately computing the current-current optical correlation function, including all the momentum-exchange processes with the crystal lattice \textcolor{black}{(refer Supplementary for benchmarks)}.

\textit{Spin-resolved deep-microscopic optical bandstructure--} A central contribution in this Letter is the spin-resolved deep-microscopic optical bandstructure of Te, presented in Figure 1(f). Inspired by the dynamics of electron waves and their corresponding bandstructure (Figure \ref{fig1}(e)), the deep-microscopic optical bandstructure (Figure \ref{fig1}(f)) unravel allowed optical indicators and corresponding hidden optical waves within Te lattice. 

For a given $\bm q$, we define the transverse optical indicators $\alpha^T_{m,\bm{q},\omega}$ as the eigenvalues of transverse symmetry-indicating lattice conductivity (SILC) tensor $\bm{\sigma}^T_{G,G^\prime}(\bm{q},\omega)$ \cite{Bharadwaj2024}:

\begin{subequations}
\begin{align}
&\bm{\sigma}^T_{\bm G, \bm G^\prime}(\bm{q},\omega)=-i\omega \varepsilon_0\sum_{\bm{m}} \frac{\alpha^T_{m,\bm{q},\omega}-1}{4\pi} \tilde{\bm{e}}_{m,\omega,\bm{q}} \otimes \tilde{\bm{e}}_{m,\omega,\bm{q}} \\
&\bm{\sigma}^T_{\bm G, \bm G^\prime}(\bm{q},\omega) \tilde{\bm{e}}_{m,\omega,\bm{q}} = -i\omega \varepsilon_0\frac{\alpha^T_{m,\bm{q},\omega}-1}{4\pi}\tilde{\bm{e}}_{m,\omega,\bm{q}},
\end{align}
\end{subequations}

\noindent where $m$ is the index of the optical band, $\varepsilon_0$ is the permittivity in free space, $\tilde{\bm{e}}_{m,\omega,\bm{q}}=[e_{m,\omega,\bm{q+G}_1},e_{m,\omega,\bm{q+G}_2}, ...]$ is the eigenvector with $e_{m,\omega,\bm{q+G}}$  as their plane-wave components, $\bm G, \bm G^\prime$ are the reciprocal lattice vectors, and the crystal momentum $\bm q$ is restricted to the first Brillouin zone. The transverse SILC tensor $\bm{\sigma}^T_{G,G^\prime}(\bm{q},\omega)$ encodes the current-current correlations including sub-nanoscale momentum transfer processes within the Te crystal lattice (Supplementary). \textcolor{black}{We note that our theory of deep-microscopic optical bandstructure is universally applicable to all crystalline systems.}

Near the $\bm \Gamma$ point, the photon momentum $\bm q\rightarrow 0$ reproduces the classical limit, where we observe that $\alpha^T_{m=1,2,\bm \Gamma,\omega}=n_o^2$, and $n_o$ being the ordinary refractive index of Te. Hence, the set of optical indicators we introduced reduces to the conventional refractive index at the $\bm \Gamma$ point. The optical indicators defined in the equations above provide a quantum generalization of the concept of refractive index. 

The plane-wave components of the eigenvectors $\tilde{\bm{e}}_{m,\omega,\bm{q}}$ provide information on the polarization texture corresponding to each point in the deep-microscopic optical bandstructure. We quantify this polarization texture by calculating the degree of circular polarization (DoCP) along the Te screw axis $\hat{\bm n}_{OA}$ (Figure \ref{fig1}(f)).  The intensity of the red/blue color in the colorbar indicates the degree of right/left circular polarization of the corresponding eigenwaves in the optical bandstructure.

\textit{Super-dispersive optical gyrotropy--} In this section, we show that the optical bandstructure captures the dispersive nature of the optical gyrotropy. Optical gyrotropy refers to the phenomenon of polarization rotation as it traverses through the crystal. It has been a long-standing challenge in the optics community to accurately predict the optical gyrotropy of undoped Te, which we address using our deep-microscopic optical band theory.

An incident photon with finite momentum along screw-axis of Te lifts the degeneracy of the optical bands (Figure \ref{fig1}(f)). For small $\bm q$, this degeneracy lifting of the dominant optical bands ($m=1,2$) is proportional to the optical gyrotropy ($\bm g_{33}\cdot \bm q_z$), given by:
\begin{align}
    \bm g_{33}\cdot \bm q_z=\frac{1}{2}\left(\alpha^T_{m=1,\bm{q},\omega}-\alpha^T_{m=2,\bm{q},\omega} \right), 
\end{align}
\noindent where $\bm g_{33}$ is the component of the optical gyrotropy tensor along the screw axis. Our theoretical plot of the optical gyrotropy in Figure \ref{fig4}(b) show excellent agreement with the experimental values for optical gyrotropy in Figure \ref{fig4}(a), providing strong validation for our theory. For the experimental values, we use the relation from \cite{Shalygin2012}: $\bm g_{33}(\omega) \cdot \bm q_z = \Theta \cdot 2cn_o/\omega$, where $\Theta$ is the optical rotatory power (optical rotation per unit length) measured from experiments, and $n_o$ is the ordinary refractive index of Te at $\omega$ \cite{Sherman1970,Bhar1976}. \textcolor{black}{We note that although we extract the gyration tensor from the small $q$ limit of the full, nonlocal current-current correlation function; however, the spin-polarized optical indicators, are calculated for all momenta across the first Brillouin zone as shown in Figure 1(f).}

The optical gyrotropy in Te display a strong wavelength dependence ($\sim 1/\lambda^2$) in the transparent mid-infrared region (Figure \ref{fig4}(a,b)), which we refer to as super-dispersive. We highlight that this super-dispersive optical gyrotropy naturally emerges from the deep-microscopic optical bandstructure, without requiring phenomenological arguments such as the coupled-oscillator model \cite{Chandrasekhar1953}.

Furthermore, optical gyrotropy is connected to the degree of chirality along the direction of photon momentum. To quantify it in the small momentum limit, we plot the chirality density on the isofrequency surface for the two dominant bands in Figure \ref{fig4}(c,d). We obtain the chirality density ($\chi$) from the expression \cite{Wang2024}: $\chi=\bar{g}_{ij} q_js_j$, where $\bar{g}_{ij} =\partial_\omega(\omega g_{ij})$, and $\bm s$ is the optical spin density in momentum space. It reflects the extent of chiral asymmetry along a particular axis. Figure \ref{fig4}(c,d) shows the highest chirality density at the poles, proving that light will encounter maximum optical gyrotropy along the $\bm \Gamma -\mathbf{A}$ axis.

\textit{Hidden optical waves--} Analogous to the electron waves which encompasses the allowed electronic modes in a material, each point on the deep-microscopic optical bandstructure corresponds to an allowed propagating optical mode within Te. In real space, the hidden optical waves are written in terms of the plane-wave components of the transverse SILC eigenvectors (${e}_{m,\omega,\bm{q},\bm G}$) as  \cite{Bharadwaj2024,Mun2024}:
\textcolor{black}{
\begin{align}
         \vec{\bm{e}}_{m,\bm q}(\bm{r},\omega,t) = \sum_{\bm G}{e}_{m,\omega,\bm{q},\bm G} \boldsymbol{n}^\perp_{\boldsymbol{q+G}} e^{i((\bm{q+G})\cdot \bm r - \omega t)},
\end{align}}
where  $\boldsymbol{n}^\perp$ is the transverse unit vector, such that $\boldsymbol{n}^\perp_{\boldsymbol{q+G}}\cdot (\bm G+\bm q)=0$. Any arbitrary electric field inside Te exists as a superposition of these optical waves over different optical bands.
The Movie 1 in (Supplementary) visualizes the propagation of optical waves for the first optical band along the Te screw axis. From Figure \ref{fig2} and the Movie 1 in (Supplementary), it is evident that these hidden waves possess a rich polarization texture at the lattice level.

\textit{Deep-microscopic optical spin texture--} To quantify the polarization texture of these hidden optical waves, we write the optical spin density within Te crystal lattice as \cite{Yang2022}:
\begin{equation}
    \vec{\bm s}^{\bm{e}_m}_{\bm q}(\bm r,\omega)=\varepsilon_0 \vec{e}_{m,\bm q}(\bm r,\omega)\times\vec{a}_{m,\bm q}(\bm r,\omega),
\end{equation}
where $\vec{a}_{m,\bm q}(\bm r,\omega)$ is the deep-microscopic vector potential corresponding to the optical wave $\vec{e}_{m,\bm q}(\bm r,\omega)$. We define the deep-microscopic optical spin texture as the spatial distribution of this optical spin density. In Figure \ref{fig3}, we present the deep-microscopic optical spin texture along the (a) $x=0$, (b) $z=0$, and (c) $y=0$ planes inside the Te crystal lattice for the first optical band. From Movie 2 in (Supplementary), we find that the optical spin vectors of the hidden optical waves rotate along the Te screw axis ($c=5.929\mathring{A}$), within the crystal lattice.

Neumann's principle suggests that a crystal that is invariant under certain symmetry, any of its physical properties must also be invariant under the same symmetry \cite{KellyGrovesKidd2000}. The deep-microscopic optical spin texture embodies the chirality of Te, hence it is consistent with this principle. Along the non-centrosymmetric axis ($\bm \Gamma\rightarrow \mathbf{A}$), the optical spin texture of the optical bands is typically higher than other points in the Brillouin zone, with the peak value near $\bm q\rightarrow 0$, evident from the DoCP. Normal to the $\bm \Gamma\rightarrow \mathbf{A}$ direction ($\bm \Gamma\rightarrow \mathbf{K}$), the optical spin texture diminishes as we move towards the $\bm \Gamma$ point, completely vanishing at $\bm q=0$.

\textcolor{black}{The deep-microscopic optical spin texture is a fundamental, lattice-scale quantity whose spatial average is directly related to the measurable DoCP of the hidden optical wave. The differential response of Te to varied DoCP manifests as optical gyrotropy, which serves as the key macroscopic observable of the microscopic spin texture. Although experimentally resolving this sub-lattice-level texture is challenging with the current state-of-the-art methods, we propose its direct observation using Spin-resolved Near-field Scanning Optical Microscopy (s-NSOM). This technique uses an AFM tip to scan the Te surface, interacting with and scattering the evanescent hidden optical waves. The scattered light can then be analyzed for its constituent Stokes parameters to reconstruct the full optical spin texture.}

\textit{Electromagnetic Lagrangian density--} So far, we have established the existence of optical waves, and showed that they contain a deep-microscopic optical spin texture when inversion symmetry is broken. However, in order to encapsulate the fundamental dynamics between these optical waves with spin texture and the Te crystal, we derive the full electromagnetic Lagrangian density.

For simplicity, we restrict ourselves to the semi-classical limit (small $\bm q$). In this limit, we can expand the dielectric permittivity as: $\varepsilon_{ij}(\omega,\bm{q})=\varepsilon^0_{ij}(\omega)+\beta_{ijk}(\omega)q_k+i\gamma_{ijk}(\omega)q_k$, where $\beta$ is purely symmetric and $\gamma$ is purely anti-symmetric. Since the chiral nature of Te allows for time-reversal symmetry, but breaks the inversion symmetry, we get $\beta=0$. We can expand the antisymmetric tensor $\gamma$ in terms of the optical gyration tensor by the relation $\gamma_{ijk}=\epsilon_{ijl}g_{lk}$, where $\epsilon$ is the Levi-Civita tensor \cite{Shalygin2012}. Furthermore, as previously mentioned, any arbitrary electric field $\bm E(\bm r,\omega)$ inside Te can be expanded in the basis of the optical waves. Thus, without loss of generality, we write the electric field as $\bm E(\bm r,\omega) = \sum_{m}c_m\vec{\bm e}_{m,\bm q}(\bm r,\omega)$, where $c_m$ are the optical wave coefficients, and are dependent on the polarization of the incoming electric field.
 
We thus obtain the full electromagnetic Lagrangian density as (Supplementary):
 \begin{widetext}
     \begin{align}
     \label{lagrangian}
    \mathcal{L}=\sum_{m,m^\prime}c_m c_{m^\prime}\Bigg(\frac{1}{8 \pi}\left(\varepsilon^0_{ij}(\omega) (\vec{\bm e}_{m,\bm q})_i (\vec{\bm e}_{m^\prime,\bm q})_j-{\vec{\bm b}}_{m,\bm q}\cdot {\vec{\bm b}}_{m^\prime,\bm q}\right)-\frac{g_{i j}}{8 \pi c} \Big(\delta_{i j} \vec{\bm e}_{m,\bm q}\cdot \dot{\vec{\bm b}}_{m^\prime,\bm q}-2 (\dot{\vec{\bm b}}_{m,\bm q})_i (\vec{\bm e}_{m^\prime,\bm q})_j\Big)\Bigg).
\end{align}
 \end{widetext}
Here, $\delta_{ij}$ is the Kronecker delta function, and ${\vec{\bm b}}_{m,\bm q}$ is the magnetic field corresponding to the hidden optical waves. We dropped the dependence of each term on $(\bm r,\omega)$ for brevity.

It is interesting to check if the electromagnetic Lagrangian contains interesting terms which couple $\bm{E}$ and $\bm B$ similar to axion electrodynamics. The lack of inversion symmetry in Te is captured by the coupling term $\bm E \cdot \dot{\bm{B}}$ between the electric field ($\bm E$) and magnetic field ($\bm B$). Note $\bm E \cdot \dot{\bm{B}}$ flips its sign under inversion.  In Te, this field coupling is explicit in the second term within the EM Lagrangian of Eq. \ref{lagrangian}. We observe that this unique coupling term is accompanied by the optical gyration tensor $g_{ij}$.

In conclusion, we developed a spin-resolved deep-microscopic optical band theory for Te. This theory uncovers a unique optical spin texture intrinsic to the Te lattice, originating from lattice symmetries. Crucially, our framework accurately predicts the super-dispersive optical gyrotropy of Te, with predictions aligning exceptionally well with experimental data. This work advances our understanding of non-local optical responses at the true lattice scale. Furthermore, our quantum theory can be expanded to understand the optical responses in emerging bounded optical materials, such as Te nanowires \cite{Medeiros2017} and 2-D tellurene \cite{Niu2023}.

\textcolor{black}{The deep-microscopic optical modes with their rich spin textures offer potential avenues for dense information encoding and transfer, particularly for designing on-chip photonic waveguides. The optical gyrotropy we model is a finite-momentum effect fundamentally linked to the nonzero Berry curvature of the Bloch states \cite{Zhong2015}. While this topological origin exists in reciprocal space, our theory reveals the real-space manifestation of these optical waves and their spin at the sub-lattice level. Future work is directed towards capturing the real-space signatures of such topological features, for instance, the optical spin texture near one of the Weyl points of Te.}

\textit{Acknowledgements--} This work was supported by the Office of Naval Research (ONR) under the award number N00014231270. J.M. acknowledges the Presidential Sejong Science fellowship (RS-2023-00252778) funded by MSIT of the Korean government. J.R. acknowledges the POSCO-POSTECH-RIST Convergence Research Center program funded by POSCO, and the National Research Foundation (NRF) grant (RS-2024-00356928) funded by MSIT of the Korean government.

\appendix

\textit{Structural properties of Te--} Te crystal is composed of helical chains of covalently bonded Te atoms along $\hat{n}_{OA}$, as shown in Figure \ref{fig1}(a,b). Each unit cell (highlighted in yellow) contains three Te atoms, with lattice parameters $a=4.457\mathring{A}$ and $c=5.929\mathring{A}$. The helical chains are linked through weak van der Waals interactions. They have two different orientations or handedness, and these enantiomers are classified as the $P3_121$ (right-handed) and $P3_221$ (left-handed) space groups (Figure \ref{fig1}(c)). in this Letter, we primarily focus on the $P3_121$ space group. \textcolor{black}{We note that all chirality-dependent observables are perfectly inverted for $P3_2 21$ with respect to $P3_1 21$. For example, optical gyrotropy and the optical spin texture for the $P3_221$ structure is identical to the $P3_121$ structure but with the opposite sign.}
\smallskip

\bibliography{apssamp}

\end{document}